\documentclass[journal,comsoc,twoside]{IEEEtran}
\usepackage{mathrsfs}

\usepackage[T1]{fontenc}% optional T1 font encoding
\usepackage[nolist]{acronym}
\usepackage{bbm}
\usepackage{graphicx}
\usepackage{booktabs}
\usepackage{xcolor}
\usepackage[normalem]{ulem}

\ifCLASSINFOpdf
\else
\fi
\usepackage{amsthm}

\newtheoremstyle{note}% <name>
{8pt}% <Space above>
{8pt}% <Space below>
{}% <Body font>
{\parindent}% <Indent amount>
{\itshape}% <Theorem head font>
{.}% <Punctuation after theorem head>
{.9em}% <Space after theorem headi>
{}% <Theorem head spec (can be left empty, meaning `normal')>

\theoremstyle{note}

\usepackage{amsmath}
\usepackage[cmintegrals]{newtxmath}
\newcommand{\insertTheTitle}{ Short Non-Binary Low-Density Parity-Check Codes for Phase Noise Channels}

\newcommand{\iw}{\mathbf u}%information word
\newcommand{\iwi}[1]{ u_{#1}}
\newcommand{\cw}{\mathbf v}%codeword
\newcommand{\cwi}[1]{ v_{#1}}%codeword symbol

\newcommand{\pcw}{\mathbf c}%permuted codeword symbol
\newcommand{\pcwi}[1]{ c_{#1}}

\newcommand{\ism}{\mathbf a}%intermediate symbol mapped
\newcommand{\ismi}[1]{ a_{#1}}

\newcommand{\mcw}{\mathbf s}% diff psk modulated codeword symbol
\newcommand{\mcwi}[1]{ s_{#1}}

\newcommand{\Lde}{\mathbf L^{\mathsf E, \mathsf{det}}}
\newcommand{\Lce}{\mathbf L^{\mathsf E, \mathsf{dec}}}

\newcommand{\Lda}{\mathbf L^{\mathsf A, \mathsf{det}}}
\newcommand{\Lca}{\mathbf L^{\mathsf A, \mathsf{dec}}}

\newcommand{\Lcapp}{\mathbf L^{\mathsf {APP}, \mathsf{dec}}}

\newcommand{\Lcappi}[1]{\mathbf L_{#1}^{\mathsf {APP}, \mathsf{dec}}}

\newcommand{\rx}{\mathbf r}
\newcommand{\rxi}[1]{ r_{#1}}

\newcommand{\Ldei}[1]{ \mathbf L_{#1}^{\mathsf E, \mathsf{det}}}
\newcommand{\Lcei}[1]{ \mathbf L_{#1}^{\mathsf E, \mathsf{dec}}}

\newcommand{\Ldai}[1]{ \mathbf L_{#1}^{\mathsf A, \mathsf{det}}}
\newcommand{\Lcai}[1]{ \mathbf L_{#1}^{\mathsf A, \mathsf{dec}}}
\newcommand{\phIncVar}{x}

\newcommand{\code}{\mathcal{C}}

\newcommand{\basem}{\mathbf B}

\newcommand{\fig}[1]{{Figure~\ref{#1}}}  
       
\newcommand{\secref}[1]{Section~\ref{#1}}

\newcommand{\gf}[1]{\mathbb{F}_{#1}}  
  
\newcommand{\chalpha}{\mathcal{X}}

\newcommand{\nChUses}{D}
\makeatletter
\newcommand{\thickhline}{%
    \noalign {\ifnum 0=`}\fi \hrule height 1pt
    \futurelet \reserved@a \@xhline
}

\newcommand{\TN}{}
\newcommand{\BM}{}

\newcommand{\respRone}{}
\newcommand{\respRtwo}{}
\newcommand{\respRthree}{}

\newcommand{\asLim}{\left(E_b/N_0\right)} %stands for asymptotic limit
\newcommand{\asLimCoh}{\asLim_{\mathsf{c}}}
\newcommand{\asLimNonCoh}{\asLim_{\mathsf{nc}}}

\newcommand{\threshNonCoh}{\asLimNonCoh^*}
\newcommand{\threshCoh}{\asLimCoh^*}
\newcommand{\shLimCoh}{\asLimCoh^{\mathsf{Sh}}}
\newcommand{\infoRateNonCoh}{\asLimNonCoh^{\mathsf{Sh}}}

\newtheorem{definition}{Definition}
\newtheorem{example}{Example}

\newcommand{\id}[2]{i(#1;#2)}

\newcommand{\E}[2] {\mathbb{E}_{#1}\left[{#2}\right] }
\newcommand{\RX}{\mathbf R}
\newcommand{\MCW}{\mathbf S}

\begin{document}
%
% paper title
% Titles are generally capitalized except for words such as a, an, and, as,
% at, but, by, for, in, nor, of, on, or, the, to and up, which are usually
% not capitalized unless they are the first or last word of the title.
% Linebreaks \\ can be used within to get better formatting as desired.
% Do not put math or special symbols in the title.
\begin{acronym}
\acro{AWGN}{additive white Gaussian noise}
\acro{BER}{bit error rate}
\acro{BP}{belief propagation}
\acro{BPSK}{binary phase shift keying}
\acro{CER}{codeword error rate}
\acro{CN}{check node}
\acro{DE}{density evolution}
\acro{DP}{discretized-phase}
\acro{DVB-S2}{Digital Video Broadcasting - Satellite 2}
\acro{EC}{erasure channel}
\acro{EXIT}{extrinsic information transfer}
\acro{PSK}{phase shift keying}
\acro{IRA}{irregular repeat-accumulate}
\acro{IT}{iterative}
\acro{LDPC}{low-density parity-check}
\acro{LLR}{log-likelihood ratio}
\acro{MAP}{maximum a posteriori}
\acro{MIMO}{multiple-input and multiple-output}
\acro{ML}{maximum-likelihood}
\acro{mMTC}{massive machine{-}type communications}
\acro{p.m.f.}{probability mass function}
\acro{p.d.f.}{probability density function}
\acro{PEG}{progressive edge growth}
\acro{PPM}{pulse position modulation}
\acro{PPCM}{pulse position coded modulation}
\acro{PSK}{phase-shift keying}
\acro{OOK}{on-off keying}
\acro{RA}{repeat-accumulate}
\acro{RCB}{random coding bound}
\acro{RS}{Reed-Solomon}
\acro{r.v.}{random variable}
\acro{SCPPM}{serially concatenated pulse position modulation}
\acro{VN}{variable node}
\acro{SPA}{sum-product algorithm}
\acro{FG}{factor graph}
\acro{PSD}{power spectral density}
\acro{DPSK}{differential phase-shift keying}
\acro{APP}{a posteriori}
\acro{SCCC}{serially concatenated convolutional code}
\acro{QPSK}{quaternary phase shift keying}
\acro{BCJR}{Bahl Cocke Jelinek Raviv}
\acro{LDGM}{low-density generator matrix}
\acro{DT}{dependency testing}
\acro{IoT}{Internet of Things}
\acro{5G}{fifth generation}
\acro{ITU}{International Telecommunication Union}
\acro{FEC}{forward error correction}
\acro{QC}{quasi-cyclic}
\end{acronym}
\title{\insertTheTitle}
%
%
% author names and IEEE memberships
% note positions of commas and nonbreaking spaces ( ~ ) LaTeX will not break
% a structure at a ~ so this keeps an author's name from being broken across
% two lines.
% use \thanks{} to gain access to the first footnote area
% a separate \thanks must be used for each paragraph as LaTeX2e's \thanks
% was not built to handle multiple paragraphs
%
\author{Tudor~Ninacs,~\IEEEmembership{Student Member,~IEEE},
		Bal\'azs~Matuz,~\IEEEmembership{Member,~IEEE},
		Gianluigi~Liva,~\IEEEmembership{Senior~Member,~IEEE},
		Giulio~Colavolpe~\IEEEmembership{Senior~Member,~IEEE}% <-this % stops a space
\thanks{
	T. Ninacs was with the the Institute of Communications and
	Navigation of the German Aerospace Center (DLR), M\"unchner Strasse 20, 82234 We{\ss}ling, Germany and is now with Intel Germany GmbH, Am Campeon 12\respRtwo{,} 85622, Munich, Germany.
	Email:tudor.ninacs@intel.com.}% <-this % stops a space		
\thanks{B. Matuz and G. Liva are with the the Institute of Communications and
	Navigation of the German Aerospace Center (DLR), M\"unchner Strasse 20, 82234 We{\ss}ling, Germany.
	Email:\{balazs.matuz, gianluigi.liva\}@dlr.de.}% <-this % stops a space
\thanks{G. Colavolpe is with the University of Parma, Department of Engineering and Architecture, and with Consorzio Nazionale Interuniversitario per le Telecomunicazioni (CNIT), viale delle scienze 181/A, 43124 Parma, Italy. Email: giulio.colavolpe@unipr.it.}
\thanks{A preliminary version of this work titled "Non-binary LDPC
	coded DPSK modulation for phase noise channels" was presented in part at the IEEE International Conference on Communications 2017, Paris, France.}
\thanks{\textcopyright 2019 IEEE.  Personal use of this material is permitted.  Permission from IEEE must be obtained for all other uses, in any current or future media, including reprinting/republishing this material for advertising or promotional purposes, creating new collective works, for resale or redistribution to servers or lists, or reuse of any copyrighted component of this work in other works. Article DOI: 10.1109/TCOMM.2019.2909201} }% <-this % stops a space

% note the % following the last \IEEEmembership and also \thanks - 
% these prevent an unwanted space from occurring between the last author name
% and the end of the author line. i.e., if you had this:
% 
% \author{....lastname \thanks{...} \thanks{...} }
%                     ^------------^------------^----Do not want these spaces!
%
% a space would be appended to the last name and could cause every name on that
% line to be shifted left slightly. This is one of those "LaTeX things". For
% instance, "\textbf{A} \textbf{B}" will typeset as "A B" not "AB". To get
% "AB" then you have to do: "\textbf{A}\textbf{B}"
% \thanks is no different in this regard, so shield the last } of each \thanks
% that ends a line with a % and do not let a space in before the next \thanks.
% Spaces after \IEEEmembership other than the last one are OK (and needed) as
% you are supposed to have spaces between the names. For what it is worth,
% this is a minor point as most people would not even notice if the said evil
% space somehow managed to creep in.

% The paper headers
\markboth{IEEE Transactions on Communications }%
{Ninacs \MakeLowercase{\textit{et al.}}: \insertTheTitle}
% The only time the second header will appear is for the odd numbered pages
% after the title page when using the twoside option.
% 
% *** Note that you probably will NOT want to include the author's ***
% *** name in the headers of peer review papers.                   ***
% You can use \ifCLASSOPTIONpeerreview for conditional compilation here if
% you desire.

% If you want to put a publisher's ID mark on the page you can do it like
% this:
%\IEEEpubid{0000--0000/00\$00.00~\copyright~2015 IEEE}
% Remember, if you use this you must call \IEEEpubidadjcol in the second
% column for its text to clear the IEEEpubid mark.

% use for special paper notices
%\IEEEspecialpapernotice{(Invited Paper)}

% make the title area
\maketitle

% As a general rule, do not put math, special symbols or citations
% in the abstract or keywords.
\begin{abstract}
This work considers the design of short non-binary \ac{LDPC} codes over finite fields of order $m$, for channels with phase noise. In particular, $m$-ary \ac{DPSK} modulated code symbols are transmitted over an \ac{AWGN} channel with Wiener phase noise. At the receiver side, non-coherent detection takes place, with the help of a multi-symbol detection algorithm, followed by a non-binary decoding step. Both the detector and decoder operate on a joint factor graph. As a benchmark, finite length bounds and information rate expressions are computed and compared with the \ac{CER} performance, as well as the iterative threshold of the obtained codes. As a result, performance within $1.2$~dB from finite-length bounds is obtained, down to a \ac{CER} of $10^{-3}$.
\end{abstract}

% Note that keywords are not normally used for peerreview papers.

% For peer review papers, you can put extra information on the cover
% page as needed:
% \ifCLASSOPTIONpeerreview
% \begin{center} \bfseries EDICS Category: 3-BBND \end{center}
% \fi
%
% For peerreview papers, this IEEEtran command inserts a page break and
% creates the second title. It will be ignored for other modes.
\IEEEpeerreviewmaketitle

\acresetall
\section{Introduction}
In the context of the upcoming {\ac{5G}} standard for cellular communications, \ac{mMTC} \TN{are} considered to be one of the key applications  {\cite{ITU-R-M.2083-0}, \cite{OBB+14}}. In this scenario, small devices, for instance sensors, sparsely transmit small amounts of data. To keep the cost of such devices small, low-end oscillators might be \TN{used}, which give rise to phase noise. Furthermore, {non-binary modulation schemes might be employed}, in order to efficiently exploit the available spectrum. {{Also,} the number of pilots for estimating the channel {is} chosen such that the overall transmission overhead is kept as small as possible}, while maintaining sufficient quality of the channel estimate \cite{DKP16}.

{Whenever short frames are considered, e.g., in the order of a few hundred symbols, pilots may yield a non-negligible loss in spectral efficiency. A remedy consists in dropping the usage of pilots and using a differential modulation scheme, such as \ac{DPSK}}\TN{,} with non-coherent detection at the receiver \cite{DS90, CBC05, Col13}. To recover {the} performance gap {with respect to} the coherent case, i.e., when full phase information is available at the receiver, non-coherent detectors, which use multiple symbols to compute a decision, may be used in practice. For sufficiently long sequences{,} they are shown to perform {close} to coherent schemes \cite{DS90}, \cite{CR00}. 

Depending on various constraints, two approaches can be taken to reliably communicate in this scenario \cite{Col13}. In {the} first approach, differential modulation can be used together with a standard  forward error correcting code \cite{PSG00}. This results in a serial turbo scheme that is then decoded by iteratively exchanging soft information between {the} detector and decoder{. This } has been previously used on a variety of channels \cite{HL99,PSG00,CBC05}. Alternatively, the channel code itself may be modified and made resilient to phase uncertainties, as demonstrated, e.g., in \cite{KC08,MLP+13}.  

{Code design for phase noise channels has been widely addressed in \TN{the} literature. In \cite {Col13,BC07} the authors investigate different detection algorithms to counteract phase noise. The detector is concatenated with \TN{the decoder of} various {binary} codes from the literature \TN{to form} a turbo \TN{detection} scheme. {Binary \ac{LDPC} code design for continuous phase frequency shift keying modulation and a blockwise non-coherent \ac{AWGN} channel was performed in \cite{PPT+17} for a bit-interleaved coded modulation scheme.} In \cite{FFR+05}, a code design for binary codes using differential modulation {was} considered. It {was shown} that taking into account the differential modulator in the code design yields performance gains. The work in \cite {BC11} extends \cite {BC07}\TN{,} by introducing an accumulator based \ac{LDPC} code design. Iterative decoding thresholds for irregular ensembles are provided\TN{,} while finite-length designs {were not investigated}. In \cite{tK03}{,} a similar scheme for multiple-input and multiple-output communications was presented, where the detector was merged with the \ac{CN} decoder of a {binary} repeat accumulate code. }

{\respRtwo{Initial} work on non-binary convolutional codes over rings\TN{,} using \ac{PSK} modulation\TN{,} date\respRtwo{s} back to \cite{MM89,FF91}, where various convolutional code designs were presented for the \ac{AWGN} channel. In order to make the codes robust against block-wise phase noise, an additional differential modulator is suggested in \cite{FF91}, without yet considering powerful turbo detection at the receiver.} 

{A binary \ac{LDPC} code design for \ac{BPSK} and Wiener phase noise\TN{,} with turbo and blind phase estimation\TN{,} \respRtwo{was} presented in \cite{KCP08}. \TN{During} code construction\TN{,} some local \acp{CN} are introduced to resolve phase ambiguities. In  \cite{KC08}, the work is extended to \ac{QPSK} using $4$-ary codes over rings. The scheme is shown to handle Wiener phase noise with {a} standard deviation of up to  $2^\circ$. In both cases\TN{,} codewords of a few thousand bits are considered. \ac{LDPC} codes over rings for \ac{PSK} modulation and the coherent \ac{AWGN} channel {were} studied in \cite{SF05}.} 
	
{In \cite{NML17+}\TN{,} a surrogate non-binary \ac{LDPC} code design over a finite field for the \ac{AWGN} channel was presented. The codes were adapted to the non-coherent phase noise channel and showed excellent performance for short blocks.} {This work is a continuation of \cite{NML17+}, where we further elaborate on the code design.}  

{\TN{In the following, we} focus on transmission of \respRtwo{short blocks} over \ac{AWGN} channels \respRtwo{with}  Wiener phase noise.} To achieve reliable communication, we make use of a coded modulation system, where a non-binary \ac{LDPC} {code} over a field of order $m$ is interfaced with a \ac{DPSK} scheme of order $m$ through a symbol interleaver. At the receiver, detection and decoding are performed on a joint factor graph\TN{,} making use of the \ac{DP} algorithm for the detector \cite{BC07} and the non-binary \ac{BP} algorithm {\cite{DM98}} for the \ac{LDPC} {code} decoder. 

This contribution differs from the literature, as the focus i\TN{s} on short blocks (in the order of a few hundred symbols) with application to \ac{mMTC}. In contrast to many \TN{existing} works, we make use of { non-binary \ac{LDPC} codes over {finite fields}\TN{,} owing to their excellent performance  over the \ac{AWGN} channel for short blocks \cite{DM98,CDD11,CML+12, PFD08}}. \respRthree{Compared} to \cite{NML17+}, we directly perform the code design of the concatenated scheme for the non-coherent Wiener phase noise channel and also present useful finite-length benchmarks for \respRthree{this} channel. \respRthree{Furthermore, we introduce a refinement step in the code design process, aiming at lowering the error-floor.}

The paper is organized as follows. \secref{sec:preliminaries} provides some background on the notation used, the channel model and the receiver structure. In \secref{sec:performance_bounds} the performance bounds used to benchmark our results are presented, followed by  \secref{sec:code_design} where the code design is described. Finally, in \secref{sec:results} some numerical results are provided and are followed by \secref{sec:conclusions} where some conclusions are drawn.

% Introduction
%%%%%%%%%%%%%%%%%%%%%%%%%%%%%%%%%%%%%%%%%%%%%%%%%%%%%%%%%%%%%%%%%%%%%%%%%%%%%%%%%%%%%%%%%%%%%%%%%%%%%%%%%%%%%%%%%%%%%%%%%%%%%%%
\section{System Setup}
\label{sec:preliminaries}
%\subsection{General Coding Scheme}

%%%%%%%%%%%%%%%%%%%%%%%%%%%%%%%%%%%%%%%%%%%%%%%%%%%%%%%%%%%%%%%%%%%%%%%%%%%%%%%%%%%%%%%%%%%%%%%%%%
\subsection{Transmitter Description}
\begin{figure}[t]
	\centering
	\includegraphics[scale=0.9, draft=false]{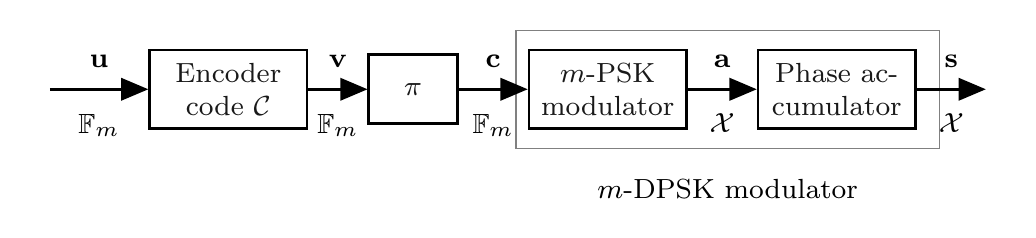}
	\caption{Transmitter block diagram.}
	\label{fig:encoder0}
\end{figure}

Throughout this paper, we will consider a coded modulation system as depicted in \fig{fig:encoder0}. Here, {a length-$K$ information frame $\iw = (\iwi{1},\iwi{2},\ldots,\iwi{K})$}, is encoded by a non-binary code $\code$ over the finite field of order $m$, $\gf{m}$. This yields {a} {length-$N$ codeword $\cw=(\cwi{1}, \cwi{2},\ldots, \cwi{N})$}. Both $\iw$ and $\cw$ are non-binary vectors whose elements belong to $\gf{m}$. 

The symbols of the codeword vector $\cw$ are then interleaved by means of a (random) interleaver $\pi$, yielding {$\pcw = (\pcwi{1}, \pcwi{2}, \ldots, \pcwi{N})$}, and input to an $m$-ary \ac{DPSK} modulator, where the field and modulation order are matched to each other. Differential modulation is performed in two steps. At first the non-binary symbols $\pcwi{i}$, are mapped to complex constellation points belonging to $\chalpha=\left\{e^{j2\pi l /m}\right\}$, $l\in\{0,\ldots,m-1\}$.\footnote{Examples of such mappings are given in \secref{sec:results}} This results in {$N$} complex modulation symbols\TN{,} {$\ism = (\ismi{1}, \ismi{2}, \ldots, \ismi{N})$, $\ismi{i}=e^{j \varphi_i}$.} In {the} second step, the phase of these symbols is accumulated, {obtaining the transmitted symbols $\mcw=(\mcwi{0}, \mcwi{1}, \ldots, \mcwi{N})$}. {By expressing $\phi_i=\arg(\mcwi{i})$, the phase accumulator implements} {$\phi_i =[\phi_{i-1} +\varphi_i ]_{2\pi}$, where $[\cdot ]_{2\pi}$ denotes the operation modulo $2\pi$}, and outputs $N + 1$ symbols, with $\mcwi{0} = 1$.  

{In the following,} we will always assume that the non-binary code order is matched to the modulation order and that $m > 2$. We also denote by $k$ and $n$ the number of information and codeword bits  in $\iw$ and $\cw$ respectively, with $k = K\log_2m$, $n = N\log_2m$. We define the code rate \respRtwo{of the code $\mathcal{C}$ as} $R_C=K/N=k/n$.\footnote{\respRtwo{Recall that the \ac{DPSK} accumulator outputs $N + 1$ symbols, where the symbol $s_0$ is a phase reference symbol. It follows that the exact code rate of the concatenated scheme is $K/(N+1) = k/(n+\log_2m)$. The difference with respect to the code rate $R_C$ turns out to be very limited for the block lengths considered in this paper. We always refer to $R_C$ in the paper.}}

%%%%%%%%%%%%%%%%%%%%%%%%%%%%%%%%%%%%%%%%%%%%%%%%%%%%%%%%%%%%%%%%%%%%%%%%%%%%%%%%%%%%%%%%%%%%%%%%%%
\subsection{Channel Model}
The \ac{DPSK} symbols $\mcwi{i}$ are transmitted over an \ac{AWGN} channel affected by phase noise. To model the phase noise, we make use of a popular model from literature \cite{Col13}, i.e., the Wiener model. Hence, the received sample $\rxi{i}$ is given by
\begin{align}
\rxi{i} &= \mcwi{i} e^{j\theta_i}  + n_i \label{eq:rx_init}  \\
%&\oeq{1} e^{j\phi_i} e^{j\theta_i}  + n_i \nonumber \\
%&\oeq{2} e^{j\psi_i} + n_i \nonumber
&= e^{j\phi_i} e^{j\theta_i}  + n_i \nonumber \\
&= e^{j\psi_i} + n_i \label{eq:rx}
\end{align}
where $\theta_i$ is an unknown phase rotation introduced by the channel and $n_i$ are {independent} \ac{AWGN} {samples distributed as}
$$n_i \sim \mathcal{CN}\left(0, 2\sigma^2\right).$$
According to the Wiener model we have that 
 \begin{equation}
\theta_i = \theta_{i-1} + \Delta\theta_i 
\label{eq::wiener}
\end{equation}
where {$\Delta\theta_{i}$ are independent, distributed as}
$$\Delta\theta_i \sim \mathcal{N} \left(0, \sigma_{\Delta}^2\right)$$
with $\theta_0$ uniformly distributed  in $[0,2\pi)$. {The} phase of the received signal $\psi_i$ is obtained as $\psi_i = [\theta_i + \phi_i]_{2\pi}$.

As a reference, we also evaluate the performance of our system on a coherent \ac{AWGN} channel, obtained by setting $\theta_{i} = 0, \forall i$ in \eqref{eq:rx_init}.
%By using the fact that the transmitted \ac{DPSK} symbols equal  $\mcwi{i}=e^{j\phi_i}$ 

%%%%%%%%%%%%%%%%%%%%%%%%%%%%%%%%%%%%%%%%%%%%%%%%%%%%%%%%%%%%%%%%%%%%%%%%%%%%%%%%%%%%%%%%%%%%%%%%%%
%\subsection{Receiver Description}
\subsection{Iterative Detection and Decoding at the Receiver}
\label{sec:it_det_dec}

\begin{figure}[t]
	\centering	
	\includegraphics[scale=0.9, draft=false]{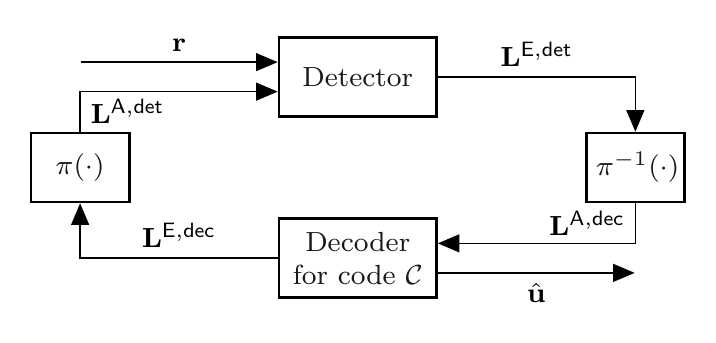}
	\caption{Block diagram of the iterative receiver.}
	\label{fig:receiver}
\end{figure}

The block diagram in \fig{fig:receiver} illustrates the exchange of messages at the receiver. First, the detector processes the received samples $\rx$ together with the a priori information $\Lda$ on the modulated codeword sequence $\ism$, available from the decoder. {The message vector $\Lda = (\Ldai{1}, \Ldai{2}, \ldots, \Ldai{N})$ is a vector of $N$ \acp{p.m.f.}, having $m$-dimensional components $\Ldai{i}$. The same holds for all the other vectors, $\Lde$, $\Lca$, $\Lce$, $\Lcapp$.} 

We have that { $\Ldai{i}=\mathbf P(\ismi{i})$} is initially set to $[1/m,\ldots,1/m]$. The detector computes soft extrinsic information $\Lde$ on the modulated codeword symbols $\ism$ with {$\Ldei{i} = k \mathbf P(\ismi{i}|\rx) / \mathbf P(\ismi{i})$, with the division performed element-wise and followed by a normalization step (denoted as such by multiplication with constant $k$)}. The elements of $\Lde$ are de-interleaved and provided as a priori information on the code symbols $\cw$ to the decoder. Second, from these a priori messages\TN{,} the decoder computes a posteriori messages $\Lcapp$\TN{,} with {$\Lcappi{i}\approx \mathbf P(\cwi{i}|\Lcai{i})$} and extrinsic messages $\Lce$\TN{,} with $\Lcei{i} ={k} \Lcappi{i} / \Lcai{i}$, {where again the division is {performed} element wise and is followed by a normalization step}. The extrinsic messages are interleaved and provided to the detector as a priori information, which can be used to compute refined estimates {of} $\Ldei{i}$. The message exchange between the decoder and detector is iterated for a certain number of times\TN{,} before a decision on the code symbols\TN{,} based on  $\Lcapp$\TN{,} is made.

In the {following}, we describe the structure of the detector based on the work in {\cite{BC07}}, \cite{NML17+}, followed by a discussion on the decoder. 

\subsubsection{Detection}

The role of the detector is to provide an estimate of the symbol-wise probability $\mathbf P(\ismi{i}|\rx)$\TN{,} {which is divided element-wise by the priors $\mathbf P(\ismi{i})$ and normalized\TN{,} to obtain the extrinsic information $\Ldei{i}$ that is forwarded to the decoder. It is computed starting with the factorization \cite{Col13} }
\begin{equation}
\label{eq::jointProb}
\begin{split}
p(\ism, \boldsymbol{\psi}|\rx) 
&= p(\rx|\ism,\boldsymbol{\psi}) p(\boldsymbol{\psi}|\ism) P(\ism) \frac{1}{p(\rx)}\\
&\propto p(\rxi{0}|\psi_0)\prod_{i = 1}^{N} p(\rxi{i}|\psi_i) p(\psi_i|\psi_{i-1}, \ismi{i}) P(\ismi{i})\,
\end{split}
\end{equation}
where due to the Wiener model and the differential modulation
$p(\psi_i|\psi_{i-1},\ldots,\psi_{0}, \ism)=p(\psi_i|\psi_{i-1},\ismi{i})$.
This factorization allows us to make use of factor graphs \cite{KFL01} and compute $P(\ismi{i}|\rx) / P(\ismi{i})$ as 

\begin{equation}
\label{eq::extrinsic}
%\begin{split}
\frac{P(\ismi{i}|\rx)}{P(\ismi{i})} =\int\displaylimits_{{0}}^{{2\pi}}\int\displaylimits_{{0}}^{{2\pi}}\alpha(\psi_{i-1})\beta(\psi_{i})p(\psi_{i}|\psi_{i-1}, \ismi{i})d\psi_id\psi_{i-1}
%\end{split}
\end{equation}
where $\alpha(\psi_i)$ and $\beta(\psi_i)$ equal \cite{BC07}
\begin{equation}
\label{eq:fwd}
\begin{split}
\alpha(\psi_i) &= p(\rxi{i}|\psi_i) \int\displaylimits_{0}^{2\pi} \left( \sum_{\ismi{i}} p(\psi_i|\psi_{i-1},\ismi{i})P(\ismi{i})\right) \alpha(\psi_{i-1}) \ d\psi_{i-1}
\end{split}
\end{equation}	
\begin{equation}
\begin{split}
\label{eq:bwd}
\beta&(\psi_i) = p(\rxi{i}|\psi_i) \int\displaylimits_{0}^{2\pi} \left( \sum_{\ismi{i+1}} p(\psi_{i+1}|, \psi_i, \ismi{i+1})P(\ismi{i+1})\right) \beta(\psi_{i+1}) \ d\psi_{i+1}
\end{split}
\end{equation}
with $\alpha(\psi_0) = p(\rxi{0}|\psi_0)$ and $\beta(\psi_N)= p(\rxi{N}|\psi_N)$.

To compute $\alpha(\psi_i)$ and $\beta(\psi_i)$, we proceed as follows. Firstly, $p(\rxi{i}|\psi_i)$ is a complex Gaussian {\ac{p.d.f.}} with mean $e^{j\psi_i}$ and variance $\sigma^2$ per dimension. {For} the coherent case, where there is no phase uncertainty, i.e., $\theta_{i} = 0, \forall i$, in the above iterations\TN{,} the probability $p(\psi_i|\psi_{i-1}, \ismi{i})$ reduces to an indicator function 
\begin{equation}
p(\psi_i|\psi_{i-1}, \ismi{i}) \overset{\theta_i = 0}{=} \mathbbm{1}(e^{j\phi_i} = \ismi{i}e^{j\phi_{i-1}}) =
\begin{cases}
1, & \text{if}\ e^{j\phi_i} = \ismi{i}e^{j\phi_{i-1}}\\
0, &\text{otherwise}
\end{cases}.
\label{eq::indicator}
\end{equation}
{The} detector implements nothing else {but} the \ac{BCJR}\cite{BCJ+74} algorithm on the trellis of the differential modulator. 

For the non-coherent case we start with the Wiener model in \eqref{eq::wiener} and the identity {$\phi_i =[\phi_{i-1} +\varphi_i ]_{2\pi}$}, which allows us to write $\psi_i = [\psi_{i-1} + \varphi_i + \Delta\theta_i]_{2\pi}$. Since $\ismi{i}=e^{j\varphi_i}$ is a deterministic mapping between $\varphi_{i}$ and $\ismi{i}$, it holds that
\begin{equation}
p(\psi_i|\psi_{i-1}, \ismi{i}) = p(\psi_i|\psi_{i-1}, \varphi_i) = p_\Delta (\psi_i - \psi_{i-1} - \varphi_{i})
\end{equation}
where $p_{\Delta}(\cdot)$ is the \ac{p.d.f.} of the phase-noise increment $\Delta\theta_i$ (modulo $2 \pi$). {For brevity we denote $\phIncVar=\Delta\theta_i$} and hence
\begin{equation}
\label{eq:pDeltaInit}
p_\Delta (\phIncVar) = \sum\displaylimits_{\ell = -\infty}^{+\infty} g(0,\sigma_{\Delta}^{2};\phIncVar - 2\pi\ell)
\end{equation}
where
\begin{equation} \label{eq:normDistrg}
g(\mu,\sigma_{\Delta}^{2};\phIncVar) = \frac{1}{\sqrt{2\pi\sigma_{\Delta}^2}}e^{-\frac{(\phIncVar - \mu)^2}{2\sigma_{\Delta}^2}}\,
\end{equation}
since the increment $\Delta\theta_i$ is normally distributed.

For the values that $\sigma_\Delta$ takes in practice, the \ac{p.d.f.} in \eqref{eq:normDistrg} is {approximately} zero in all points, except {for} some {points} in the vicinity of $\mu$ \cite{BC11}. Hence, we can approximate
\begin{equation}
p_\Delta (\phIncVar) \simeq g(0,\sigma_{\Delta}^{2};\phIncVar)
\end{equation}
and simplify \eqref{eq:pDeltaInit} to
\begin{equation} \label{eq:pDeltaFinal}
p_\Delta (\psi_i - \psi_{i-1} - \varphi_{i}) \simeq g(0,\sigma_{\Delta}^{2};\psi_i - \psi_{i-1} - \varphi_{i})\,.
\end{equation}

Still, using \eqref{eq:pDeltaFinal} in \eqref{eq::extrinsic}, \eqref{eq:fwd} and \eqref{eq:bwd}\TN{,} the computations are rather complex, since they involve evaluating integrals of continuous \acp{p.d.f.}. A possible solution to this problem is to discretize the channel phase and implement {the} so-called \ac{DP} algorithm \cite{Col13}. We hence assume that $\psi_i$ is discrete and belongs to the set $\{2\pi j/L\}$, $j\in\{ 0, \ldots, L - 1\}$, with $L$ being the number of discretization levels. Moreover, \cite{PSG00} suggests using a further simplification
\begin{equation}
p_\Delta(\phIncVar) =
\begin{cases}
1-P_\Delta, & \phIncVar=0\\
\frac{P_\Delta}{2}, &|\phIncVar|=\frac{2\pi}{L}\\
0, &\text{else}
\end{cases}
\label{eq::pDeltaDP}
\end{equation}
\respRtwo{with $P_\Delta$ being an optimization parameter obtained via simulation. For all our simulations we have used $P_\Delta=0.1$}. It has been shown that a phase discretization factor of $L = 8 m$ is enough to obtain negligible losses with respect to the unquantized case. With these two approximations, the integrals in \eqref{eq::extrinsic}, \eqref{eq:fwd} and \eqref{eq:bwd} become summations and the computation of all values above becomes feasible in practice.

\subsubsection{Decoding}
The code $\code$ is assumed to be an \ac{LDPC} code. Thus, standard belief propagation for non-binary \ac{LDPC} codes from the literature can be applied. For more details on non-binary {decoding} of \ac{LDPC} codes\TN{,} the reader is referred, e.g., to {\cite{DM98}, \cite{DF07}}. For our setup{,} we perform only one iteration of the belief propagation algorithm {within the decoder at a time, and allow a maximum of} $N_{\mathrm{it}} = 200$ iterations between {the} detector and decoder. {This value was chosen in accordance with the literature on non-binary \ac{LDPC} codes (see, e.g., \cite{DM98,CML+12,WSM04,SD10}).}

% Basics
\section{Performance Bounds}
\label{sec:performance_bounds}

%\subsection{Information Rate and DT Bound}
We use two benchmarks to \respRone{assess} the performance of our system. The first one is the information rate, which gives a lower bound on the achievable rate when the block length goes to infinity. It is defined as
\begin{align}
 \lim\limits_{N \rightarrow \infty} \frac{1}{N} \E{}{\log_2 \frac{p(\RX|\MCW)}{p_{}(\RX)}} \label{eq:IR}
= \lim\limits_{N \rightarrow \infty} \frac{1}{N} \E{}{\id{\MCW}{\RX}}
\end{align}
where $\id{\cdot}{\cdot}$ denotes the information density and $\MCW$ and $\RX$ are random vectors {associated to the process describing the transmitted and received symbols\respRtwo{,} respectively. To compute it, we resort to the methods of \cite{ALP+06} as described in \cite{BC11}.

{As a finite-length performance benchmark we compute the \ac{DT} bound \cite{PPV10}, which provides an upper bound to the average block error probability $P_B$ of a random code with $M=m^K$ codewords of length {$N+1$}.  
Following \cite{PPV10} we obtain 
\begin{align}
{P_B} & {\leq \E{}{ 2^{  - \Big( \id{\MCW}{\RX} - \log_2 \frac{M - 1}{2} \Big)^{+} } } \label{eq:DT}}\\
%& \approx \E{}{ 2^{  - \Big( \id{\MCW}{\RX} - (K \log_2m-1) \Big)^{+} } }\\
%&{ = \E{}{ 2^{  - \Big( \id{\MCW}{\RX} - (\log_2(m^K - 1) - 1) \Big)^{+} } }}\\
&{\approx \frac{1}{\nChUses} \sum_{ (\mcw,\rx)}^{}{ 2^{  - \Big( \id{\mcw}{\rx} - (K \log_2m-1) \Big)^{+} }}}
%& {\approx \frac{1}{\nChUses} \sum_{ (\mcw,\rx)}^{}{ 2^{  - \Big( \id{\mcw}{\rx} - (\log_2(m^K-1)) \Big)^{+} } }} 
\label{eq:DTapprox}
\end{align}
where  $(x)^{+} \equiv \max(x, 0)$ and $\nChUses$ is the number of $(\mcw,\rx)$ tuples.}
Analogously to the computation of the information rate, we compute the information density as described in \cite{PPV10}, following a Monte Carlo approach. To this end, we randomly generate an input sequence \TN{of \ac{DPSK} modulated symbols} $\mcw$, which we transmit over the communication channel to obtain $\rx$. {For the tuple $(\mcw,\rx)$ we then evaluate the information density $\id{\mcw}{\rx}$ and the corresponding summand in \eqref{eq:DTapprox}. We repeat this experiment  $\nChUses$ times and average over the outcomes. For the communication channel used to compute $\rx$, we either use a Wiener phase noise channel, as defined in \eqref{eq:rx} or a coherent \ac{AWGN} channel, both yielding different {expressions} for the information density in~\eqref{eq:DTapprox}.}

\section{Code Design}
\label{sec:code_design}

We are interested in the design of $m$-ary \ac{LDPC} codes for $m$-\ac{DPSK} modulation over {a} non-coherent Wiener phase noise channel\TN{,} described in Section~\ref{sec:preliminaries}. Our methodology for the code design is as follows. First, we aim to find a protograph \ac{LDPC} code ensemble with {an} iterative decoding threshold close to the theoretically achievable limit. In an optional second step, we refine the protograph code design\TN{,} aiming at error floors below a target \BM{block error probability}. Next, a brief introduction on protograph \ac{LDPC} codes is given\TN{,} followed by a discussion on the computation of the iterative decoding threshold. Then, a detailed description of the protograph search algorithm is provided. The section is complemented \TN{by} some remarks on the algorithm.

%%%%%%%%%%%%%%%%%%%%%%%%%%%%%%%%%%%%%%%%%%%%%%%%%%%%%%%%%%%%%%%%%%
\subsection{Protograph LDPC Codes} \label{sec:proto}
{Protograph-based binary} \ac{LDPC} codes were originally introduced in \cite{Tho03}. This class of structured \ac{LDPC} codes performs excellently on a wide class of communication channels while the code structure permits hardware friendly implementations. {A protograph can be any Tanner graph, typically one with a relatively small number of nodes \cite{Tho03}} which are connected by single or multiple edges. In the protograph, each \ac{VN} and \ac{CN} is said to be of a certain type. The protograph can be seen as a template for the bipartite graph of an \ac{LDPC} code\TN{,} which is obtained by lifting the protograph through ``copy-and-permute'' operations. For this, $\ell$ copies of the protograph are generated and interconnected as follows.  
Edges among all copies are permuted such that if a node of type $i$ was connected to a node of type $j$ in the protograph, then any of its $\ell$ copies are connected to any of the $\ell$ copies of the node of type $j$. After expansion{,} parallel edges are no longer permitted. In order to optimize the girth of the resulting graph, we perform the expansion by a circulant version of the \ac{PEG}  algorithm \cite{HEA01}. A protograph can be represented by a $m_b \times n_b$ base matrix $\basem$ whose entries $b_{ij}$ give the number of edges connecting a \ac{CN} of type $i$ to a \ac{VN} of type $j$.\footnote{{The expansion factor $\ell$ {can} be computed as $\ell=[N/n_b]$, where the squared brackets denote the "nearest integer" function.}} Note that a protograph, or alternatively its base matrix\TN{,} describe an ensemble of \ac{LDPC} codes. %, since through expansion different codes can be obtained from a single protograph.

Non-binary protographs were first introduced in \cite{CML+12}\TN{,} and can be divided into constrained and unconstrained protographs \cite{CDD12}. The former ones possess additional edge labels from $\gf{m} \backslash \{0\}$. After expansion, these labels correspond to the non-binary coefficients in the code's parity-check matrix. In this work, our attention is on unconstrained protographs\TN{,} for which no edge labels are assigned {at} protograph level. Rather, the edge labels are assigned after the final expansion step {and are chosen uniformly at random from $\gf{m} \backslash \{0\}$}.

%%%%%%%%%%%%%%%%%%%%%%%%%%%%%%%%%%%%%%%%%%%%%%%%%%%%%%%%%%%%%%%%%%
\subsection{Iterative Decoding Threshold Computation}
The iterative decoding threshold of an \ac{LDPC} code ensemble is defined as the worst channel parameter for which the ensemble average probability of symbol error vanishes\TN{,} when the block length and the number of decoding iterations go to infinity. Iterative decoding thresholds  of unstructured non-binary \ac{LDPC} code ensembles for \ac{AWGN} channels can be conveniently computed by making use of \ac{EXIT} analysis \cite{BB06}. The extension to non-binary protograph ensembles can be done by adapting the results in \cite{LC07}. 

We have computed iterative decoding thresholds for protograph \ac{LDPC} code ensembles over $\gf{m}$ adopting Method~1 from \cite{BB06}. Here, the log-probability ratios\TN{,} passed on the edges of the bipartite graph\TN{,} are approximated as multivariate Gaussian random variables. We {have} found empirically that the computed thresholds obtained by Method~1 provide limited accuracy for the setup in \fig{fig:encoder0}. To increase the accuracy of the threshold computation, the authors in \cite{BB06} propose {Method~2. This method can be adapted to non-binary protograph \ac{LDPC} codes and requires} measuring the transfer function\TN{,} for each \ac{VN} and \ac{CN} type in the protograph\TN{,} which relates the extrinsic mutual information at the output of a node to the a priori mutual information at its input.  Measuring the transfer function imposes a high computational burden, in particular if various protographs are tested, each with different node types. {In this case, \ac{EXIT} analysis {loses} its advantage of providing a low-complexity alternative to other techniques, such as Monte Carlo density evolution \cite{Mac02}.} 

{{We} therefore resort to Monte Carlo density evolution \cite{Mac02} to obtain the thresholds.} In brief, the iterative decoding threshold of a protograph \ac{LDPC} code ensemble  is obtained by performing decoding on a large bipartite graph\TN{,} where iteration by iteration\TN{,} the edge interleavers between the different node types are changed in order to emulate the average ensemble behavior (see \cite{Mac02,MPZ+17} for details). {We also make use of channel adapters  for the iterative decoding threshold computation and resort to the all-zero codeword assumption \cite{BB06}.} {Note that, owing to the protograph structure of the LDPC codes, we place an interleaver between the detector and decoder, similarly to \cite{BDM+98}. For the threshold computation we use a different random interleaver for every decoding attempt.} The computational cost  of Monte Carlo density evolution is still too high to enable the use of iterative optimization algorithms, such as differential evolution \cite{SS05}\TN{, for the search of protographs with good iterative decoding thresholds}. For this reason, we propose a simplified protograph search methodology\TN{,} aiming to reduce the {protograph} search space.

%%%%%%%%%%%%%%%%%%%%%%%%%%%%%%%%%%%%%%%%%%%%%%%%%%%%%%%%%%%%%%%%%%
\subsection{Protograph Search} \label{sec:prot_search}
{On the coherent \ac{AWGN} channel, }let us denote the input constrained Shannon limit {in terms of energy per information bit to noise power spectral density ratio} by $\shLimCoh$ and the iterative decoding threshold of a protograph \ac{LDPC} code ensemble by $\threshCoh$. Similarly, on  the non-coherent Wiener phase noise channel the theoretical limit from the information rate expression in Section~\ref{sec:performance_bounds} is named $\infoRateNonCoh$, while the iterative decoding threshold of a protograph ensemble is termed $\threshNonCoh$. Also, we denote by  $\mathbb Z_p$ the set of non-negative integers smaller than $p$. We introduce the following definitions.

\begin{definition}
	\BM{An $m_b \times n_b$ single entry matrix $\mathbf Q$ is a matrix whose entry $q_{i,j}=1$ for some $i,j$ and all other entries are set to zero.} %$q_{i',j'}=0$, $\forall i'\in \{0,1,\ldots, m_b-1\}\backslash \{i\}$ and  $\forall j'\in \{0,1,\ldots, n_b-1\}\backslash \{j\}$.
\end{definition}
\begin{definition}
		\BM{A minimal set $\mathcal M_{ e}$ of $m_b \times n_b$ matrices is a set for which an element {$\mathbf B \in \mathcal M_{ e}$} cannot be obtained by row and/or column permutation of any other element in $\mathcal M_{ e}$.} %{$\mathbf B_j \in \mathcal M_{ e}, \forall j\neq i$}. %Formally, for any $\mathbf B^{}, \mathbf B' \in \mathcal M_{\mathsf u}$ we require $\mathbf B^{} \neq   \mathbf P_{m_b\times m_b} \mathbf B' \mathbf P_{n_b\times n_b}$, where $\mathbf P_{i \times i}$ is an $i \times i$ permutation matrix.
\end{definition}
Minimal sets are of particular interest\TN{,} since the iterative decoding threshold of a protograph does not change by permuting the rows and/or columns of the associated base matrix. {Hence, in the {following}, we start from a set $\mathcal M$ of protograph base matrices and generate a minimal set $\mathcal M_{e}$ out of it\TN{,} as follows.} We start with an empty set $\mathcal M_{e}$ and pick one element of  $\mathcal M$ after the other.\footnote{The ordering of the elements of $\mathcal M$ does not play a role in our case.}  We include an element of $\mathcal M$ in $\mathcal M_{e}$, if\TN{,} after inclusion\TN{,} $\mathcal M_{e}$ is still a minimal set. Otherwise, the element is \BM{rejected}%not included in $\mathcal M_{e}$
. We formalize the protograph search algorithm as follows.

\bigskip

\subsubsection*{First Step (Threshold Optimization)}  
Our objective  is to find a protograph with iterative decoding threshold $\threshNonCoh$ on the Wiener phase noise channel as close as possible to $\infoRateNonCoh$. We consider only a small number \respRtwo{of} protographs for which iterative decoding thresholds are computed and proceed as follows.

{First, generate all $p^{m_b n_b}$ $m_b \times n_b$ base matrices whose elements {$b_{i,j}$} are picked from $\mathbb Z_p$ yielding the set $\mathcal M$. Expurgate $\mathcal M$ by imposing constraints on the base matrices {contained in it}: discard an element if it contains zero weight columns or if the number of weight-$1$ columns exceeds $m_b$. %\footnote{A motivation for this restriction is given in Section~\ref{sec:code_design_results}.} 
Generate a minimal set $\mathcal M_{ e}$  out of the expurgated set and compute iterative decoding thresholds for the elements of $\mathcal M_{ e}$.} Select the base matrix $\basem_{\text{}}^*$ with the best iterative decoding threshold and expand it to obtain an $(N,K)$ \ac{LDPC} code as discussed in Section~\ref{sec:proto}. Finally, \BM{evaluate the code performance on the Wiener phase noise channel by Monte Carlo simulation}.

\bigskip

\subsubsection*{Second Step (Refinement)} 
{If the simulation results show a visible error floor above a target \BM{block error probability}, we attempt to lower the error floor by changing the code design as follows. 
	
The base matrix $\basem_{\text{}}^*$ from step 1) is
expanded by a factor of \respRthree{$\ell'$}, where $\respRthree{\ell'}=\max_{i,j} b^*_{i,j}$ is the \BM{largest} base matrix entry. The expansion is done {according to the description in Section~\ref{sec:proto}}. This yields the $ m_b' \times   n_b'$ base matrix $ \basem_{\text{}}'${, with $m_b'=\respRthree{\ell'} m_b$.} \BM{Generate a new set ${ \mathcal M}'$ where each element is obtained by adding to $ \basem_{\text{}}'$ a different $ m_b' \times n_b'$ single entry matrix. This yields a set with cardinality $|{ \mathcal M}'|=m_b'n_b'$, since there are $m_b'n_b'$ distinct  $ m_b' \times   n_b'$ single entry matrices.} 
Note that the matrices in ${ \mathcal M}'$ have an increased average column and row weight {with respect to} $\basem_{\text{}}^*$\TN{,} which is \BM{expected to improve the distance properties of the corresponding ensemble and hence to lower the error floor (see, e.g., \cite{HE04,LPC13}).} Next, a minimal set ${\mathcal M}'_{ e}$ is generated out of ${ \mathcal M}'$. Iterative decoding thresholds  for the base matrices in ${\mathcal M}'_{ e}$ are computed and  the one %$\bar \basem_{\text{}}^*$ 
with the best iterative decoding threshold is selected. By expansion, an  $(N,K)$ \ac{LDPC} code is obtained %from $\bar  \basem_{\text{}}^*$ 
and simulated on the Wiener phase noise channel. {In the case \TN{that} the error floor is no longer visible above the target block error probability the algorithm stops, otherwise }\BM{step~2 is repeated by selecting the next best candidate in ${\mathcal M}_{ e}$.} 

%%%%%%%%%%%%%%%%%%%%%%%%%%%%%%%%%%%%%%%%%%%%%%%%%%%%%%%%%%%%%%%%%%

\subsection{Remarks}

{We conclude the section with the following remarks. Firstly, for a given code rate, the dimensions $m_b$ and $n_b$ of the base matrix are picked to be as small as possible in order to limit the search space. For instance, for code rates $R=(r-1)/r$\TN{,} base matrices of size $1 \times r$ are considered. Secondly, the base matrix entries $b_{i,j}$ are chosen from $ \mathbb Z_4$. This is motivated by the fact that non-binary  \ac{LDPC} codes with \ac{VN} degrees of three and less show excellent performance on Gaussian channels \cite{PFD08,CDD11}.} 

\section{Numerical Results}
\label{sec:results}

{In the following, we present some code design examples by applying the rule described in \secref{sec:code_design}}. {We also provide theoretical benchmarks based on the results in \secref{sec:performance_bounds}. In particular, for the coherent \ac{AWGN} case, {the} Shannon limit {$\shLimCoh$} and \ac{DT} bound are computed. For the non-coherent case, the respective theoretical limit $\infoRateNonCoh$ and \ac{DT} bound are given. Different \ac{DPSK} orders (thus field orders), code rates and standard deviations of {the} phase noise increment are considered.  {In particular, the standard deviation of the phase noise increment is $\sigma_{\Delta} = 2^\circ$ for 8-\ac{PSK} and $\sigma_{\Delta} = 1^\circ$ for 16-\ac{PSK}}.\footnote{Note that the chosen {values represent worse} case scenarios for the phase noise for \ac{DVB-S2} \cite{ETSI09a} or \ac{5G} \cite{3gpp_phase_noise}. This can be seen by comparing the respective phase noise masks with the \ac{PSD} of the Wiener process with $\sigma_{\Delta}=2^{\circ}$ or $\sigma_{\Delta}=1^{\circ}$.} The mapping between field elements and $8$-\ac{PSK}, as well as $16$-\ac{PSK} symbols\TN{,} \respRtwo{are} provided in Tables~\ref{tab:mapping8} and \ref{tab:mapping16}, respectively.  A target block error probability of $10^{-3}$ is assumed\TN{,} above which no visible error floor should occur. This falls in the range of error probabilities currently discussed for \ac{mMTC} in \ac{5G}.}

\begin{table}[t]
	\centering
	\caption{Mapping between $\gf{8}$ code symbols, their binary image, and Gray labeled $8$-\ac{PSK} modulation symbols. The primitive polynomial for $\gf{8}$ is $1+\mathrm x+\mathrm x^3$.}\label{tab:mapping8}
	\begin{tabular}{c|c|c}  
		\toprule
		$\gf{8}$ element & Binary label & $8$-PSK symbol\\
		\midrule		
		$0$ & $000$ & $1$ \\
		$\alpha^0$ &$001$ & $e^{j \pi/4}$\\	
		$\alpha^1$ &$010$  & $e^{j 3\pi/4}$	\\	
		$\alpha^2$ &$100$  & $e^{j 7\pi/4}$\\	
		$\alpha^3$& $011$  & $e^{j \pi/2}$\\	
		$\alpha^4$&	$110$  & $e^{j \pi}$\\				     	
		$\alpha^5$& $111$  & $e^{j 5\pi/4}$\\	
		$\alpha^6$ & $101$  & $e^{j 3\pi/2}$\\  
		\bottomrule 					
	\end{tabular}
\end{table}

\begin{table}[t]
	\centering
	\caption{Mapping between $\gf{16}$ code symbols, their binary image, and Gray labeled $16$-\ac{PSK} modulation symbols. The primitive polynomial for $\gf{16}$ is $1+\mathrm x+\mathrm x^4$.}\label{tab:mapping16}
	\begin{tabular}{c|c|c}  
		\toprule
		$\gf{16}$ element & Binary label & $16$-PSK symbol\\
		\midrule		
		$0$ 			& $0000$ & $1$ \\
		$\alpha^0	$ 	& $0001$ & $e^{j \pi/8}$ \\
		$\alpha^1	$ 	& $0010$ & $e^{j 3\pi/8}$ \\
		$\alpha^2	$ 	& $0100$ & $e^{j 7\pi/8}$ \\
		$\alpha^3	$	& $1000$ & $e^{j 15\pi/8}$ \\
		$\alpha^4	$	& $0011$ & $e^{j \pi/4}$ \\ 
		$\alpha^5	$	& $0110$ & $e^{j \pi/2}$ \\
		$\alpha^6	$ 	& $1100$ & $e^{j\pi}$ \\
		$\alpha^7	$ 	& $1011$ & $e^{j 13\pi/8}$ \\
		$\alpha^8	$ 	& $0101$ & $e^{j 3\pi/4}$ \\
		$\alpha^9	$ 	& $1010$ & $e^{j 3\pi/2}$ \\
		$\alpha^{10}$ 	& $0111$ & $e^{j 5\pi/8}$ \\
		$\alpha^{11}$	& $1110$ & $e^{j 11\pi/8}$ \\
		$\alpha^{12}$	& $1111$ & $e^{j 5\pi/4}$ \\
		$\alpha^{13}$	& $1101$ & $e^{j 9\pi/8}$ \\
		$\alpha^{14}$ 	& $1001$ & $e^{j 7\pi/4}$ \\
		\bottomrule 					
	\end{tabular}
\end{table}

\begin{example}[$R_c=1/2$, $8$-\ac{DPSK}] 
	Step 1 of the protograph-search for the Wiener phase noise channel  yields the set $\mathcal M_{e}$ of $1 \times 2$ base matrices. All elements of $\mathcal M_{e}$ are given \respRtwo{in} the upper part of Table~\ref{tab:thresh:8psk_r05}. {The Shannon limit for the coherent case is $\shLimCoh = 1.28$~ \textup dB. For the  non-coherent channel $\infoRateNonCoh = 1.56$~ \textup dB.} We find that among all {the} tested candidates the protograph with base matrix $\basem^{\mathrm{I}}_1=[2~1]$ has the best threshold $\threshNonCoh=2.11$\BM{~dB} on the non-coherent phase noise channel. We designed an $8$-ary  $(160,80)$ \ac{LDPC} code with rate $R_c=1/2$ from $\basem^{\mathrm{I}}_1$, where code parameters are given in symbols belonging to $\gf{8}$. %{and the lifting factor $\ell$ can be computed as $\ell=N/n_b$.}
	
	Simulation results for both {the} coherent and non-coherent channels in terms of \ac{CER} versus $E_b/N_0$ are given in \fig{fig:8psk_comp_turbo}. We observe that both in the coherent, as well as in the non-coherent case, the gap to the \ac{DT} bound is around $1$~dB. Since an error floor above the target \BM{block error probability of} $10^{-3}$ occurs, step 2 of the protograph search \BM{is performed}. This yields the set ${ \mathcal M}'_e$ consisting of three $2 \times 4$ base matrices given in the lower part of Table~\ref{tab:thresh:8psk_r05}. The base matrix $ {\mathbf B}^{\mathrm I}_{1,1}$ is selected  since it has the lowest threshold among all elements in ${ \mathcal M}'_e$. 
	With a minor loss in the waterfall performance, the error floor {no longer appears} in the simulated $E_b/N_0$ regime, {as can be seen in \fig{fig:8psk_comp_turbo}}. We {note} that the gap between the two \ac{DT} bounds is similar to the gap between the \ac{CER} performance for the code with base matrix ${\mathbf B}^{\mathrm{I}}_{1,1}$. This suggests robustness against phase noise, at least when $\sigma_{\Delta}$ is not larger than $2^{\circ}$.
	
	As a benchmark, we compare our scheme with a competitor from the literature. To this end, we adopt the {serially concatenated scheme from \cite{Col13} in the absence of pilots, where the detector implements the \ac{DP} algorithm. %In contrast to our setup, the detector is interfaced through an interleaver with a \emph{binary}} convolutional code with generators $(5, 7)$ in octal notation. 
 	The difference to our setup is that the outer channel code in \cite{Col13} is a binary convolutional code with generators $(5, 7)$ in octal notation. Both detector and convolutional code decoder iteratively exchange messages\TN{,} yielding  a powerful serial turbo code}, denoted as such in \fig{fig:8psk_comp_turbo}. We observe a loss of around 0.7~dB  {of the turbo code} with respect to our \ac{LDPC} protograph code for both the coherent and non-coherent case. We may further improve the error floor performance of the turbo scheme by increasing the memory of the binary convolutional code, which yields a small sacrifice in the waterfall performance \respRone{for the coherent case}. The performance of the turbo scheme having a 16 state $(23, 25)$ outer convolutional code is also depicted in \fig{fig:8psk_comp_turbo}.

\end{example}
\begin{table}[t]
	\centering
	\renewcommand{\arraystretch}{1.5}%streches the lines such that sub and superscripts do not overlap
	\caption{Iterative decoding thresholds for the non-coherent and coherent \ac{AWGN} channel for $8$-\ac{DPSK} modulation and rate-$1/2$ protographs. }\label{tab:thresh:8psk_r05}
	\begin{tabular}{c|c|c|c}  
		\toprule
		&{Base Matrix} & $\threshNonCoh$ [dB] & $\threshCoh$ [dB] \\
		\midrule		
		$\mathbf B^{\mathrm{I}}_1$ &$[2\ 1]$ &2.11& 1.84\\
		
		$\mathbf B^{\mathrm{I}}_2$ &$[3\ 1]$ &2.51& 2.35\\
		$\mathbf B^{\mathrm{I}}_3$ &$[2\ 2]$ &3.05& 2.82\\
		$\mathbf B^{\mathrm{I}}_4$ &$[3\ 2]$ &3.91& 3.66 \\	
		$\mathbf B^{\mathrm{I}}_5$ &$[3\ 3]$ &4.78& 4.46 \\
		\midrule
		$ {\mathbf B}^{\mathrm I}_{1,1}$ &		$\left[
		\begin{array}{cccc}
		2 & 1 & 1 & 0 \\
		1 & 1 & 0 & 1\\
		\end{array}
		\right]$ &2.18&1.98\\
		$ {\mathbf B}^{\mathrm I}_{1,2}$ &
		$\left[
		\begin{array}{cccc}
		1 & 1 & 2 & 0 \\
		1 & 1 & 0 & 1\\
		\end{array}
		\right]$ &2.51&2.31\\
		$ {\mathbf B}^{\mathrm I}_{1,3}$ &
		$\left[
		\begin{array}{cccc}
		1 & 1 & 1 & 0 \\
		1 & 1 & 1 & 1\\
		\end{array}
		\right]$ &2.35&2.15\\
		\bottomrule 					
	\end{tabular}
\end{table}
\begin{figure}[t]
	\centering	
	\includegraphics[scale=0.8, draft=false]{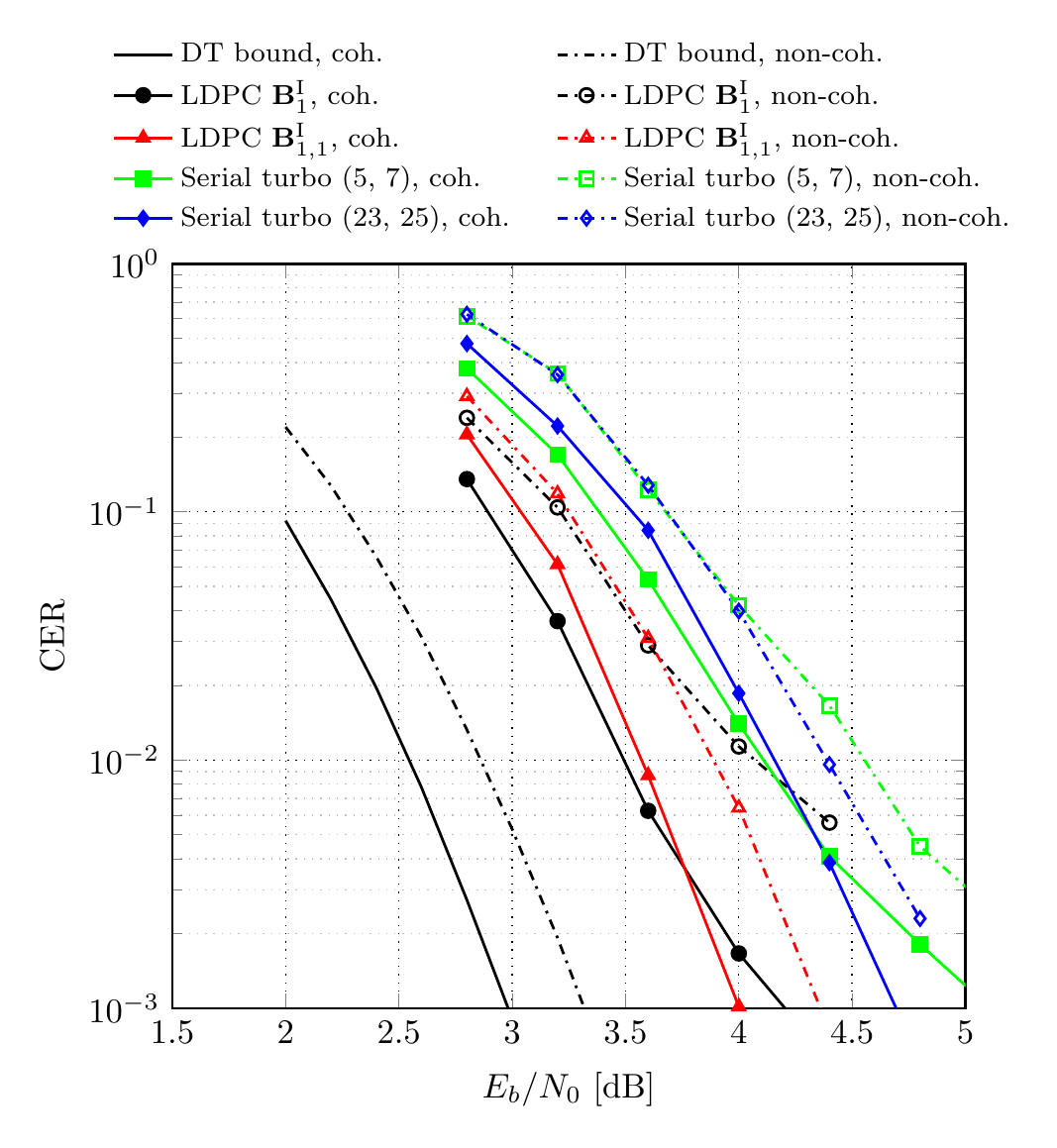}
	\caption{Comparison between {\ac{LDPC}} codes over $\gf{8}$ with base matrices $\mathbf{B}^\mathrm{I}_1$ and ${\mathbf{B}}^\mathrm{I}_{1,1}$ with a serial turbo scheme, combined with 8-\ac{PSK} modulation, code rate ${1}/{2}$, $N=160$ symbols and phase-noise having $\sigma_{\Delta} = 2^\circ$ for the non-coherent case.}
	\label{fig:8psk_comp_turbo}
\end{figure}

\begin{example}[$R_c=2/3$, $8$-\ac{DPSK}]
	Step 1 of the protograph search for the Wiener phase noise channel yields the set $\mathcal M_{e}$ of $1 \times 3$ base matrices. Iterative thresholds for all elements are given in Table~\ref{tab:thresh:8psk_r066} for both {the} coherent and non-coherent channels.  {The Shannon limit for the coherent case is $\shLimCoh = 2.76$~ \textup dB. For the  non-coherent channel $\infoRateNonCoh = 3.15$~ \textup dB.} We find that the protograph with base matrix $\basem^{\mathrm{II}}_1=[2~2~1]$ has the best threshold $\threshNonCoh=3.81$~dB {among all $7$ candidates in Table~\ref{tab:thresh:8psk_r066}}. We build out of it a rate-$2/3$ code with parameters $(120,80)$ and plot the \ac{CER} versus $E_b/N_0$ on both the coherent and non-coherent \ac{AWGN} channels in \fig{fig:8psk_0.66}. {We} observe from the figure that the gap to the \ac{DT} bounds is around 1~dB, respectively. No visible error floor is \BM{present at the target block error probability}. 
	
\end{example}
\begin{table}[t]
	\centering
	\renewcommand{\arraystretch}{1.5}
	\caption{Iterative decoding thresholds for the non-coherent and coherent \ac{AWGN} channel for $8$-\ac{DPSK} modulation and rate-$2/3$ protographs. }
	\label{tab:thresh:8psk_r066}
	\begin{tabular}{c|c|c|c}  
		\toprule
		&{Base Matrix} & $\threshNonCoh$ [dB] & $\threshCoh$ [dB] \\
		\midrule		
		%	$\mathbf B^{\mathrm{II}}_1$ &$[2\ 1\ 1]$ &\textcolor{red}{4.79} / 3.46& 3.08\\
		%	$\mathbf B^{\mathrm{II}}_3$ &$[3\ 1\ 1]$ &\textcolor{red}{4.92} / 3.41& 3.07\\
		$\mathbf B^{\mathrm{II}}_1$ &$[2\ 2\ 1]$ &3.81& 3.44\\
		$\mathbf B^{\mathrm{II}}_2$ &$[3\ 2\ 1]$ &4.15& 3.77\\
		$\mathbf B^{\mathrm{II}}_3$ &$[3\ 3\ 1]$ &4.62& 4.21\\	
		$\mathbf B^{\mathrm{II}}_4$ &$[2\ 2\ 2]$ &4.42&4.02 \\	
		$\mathbf B^{\mathrm{II}}_5$ &$[3\ 2\ 2]$ &4.80& 4.37\\	
		$\mathbf B^{\mathrm{II}}_6$ &$[3\ 3\ 2]$ &5.19&4.75 \\	
		$\mathbf B^{\mathrm{II}}_7$ &$[3\ 3\ 3]$ &5.55&5.09 \\	

		\bottomrule 					
	\end{tabular}
\end{table}
\begin{figure}[t]
	\centering	
	\includegraphics[scale=0.8, draft=false]{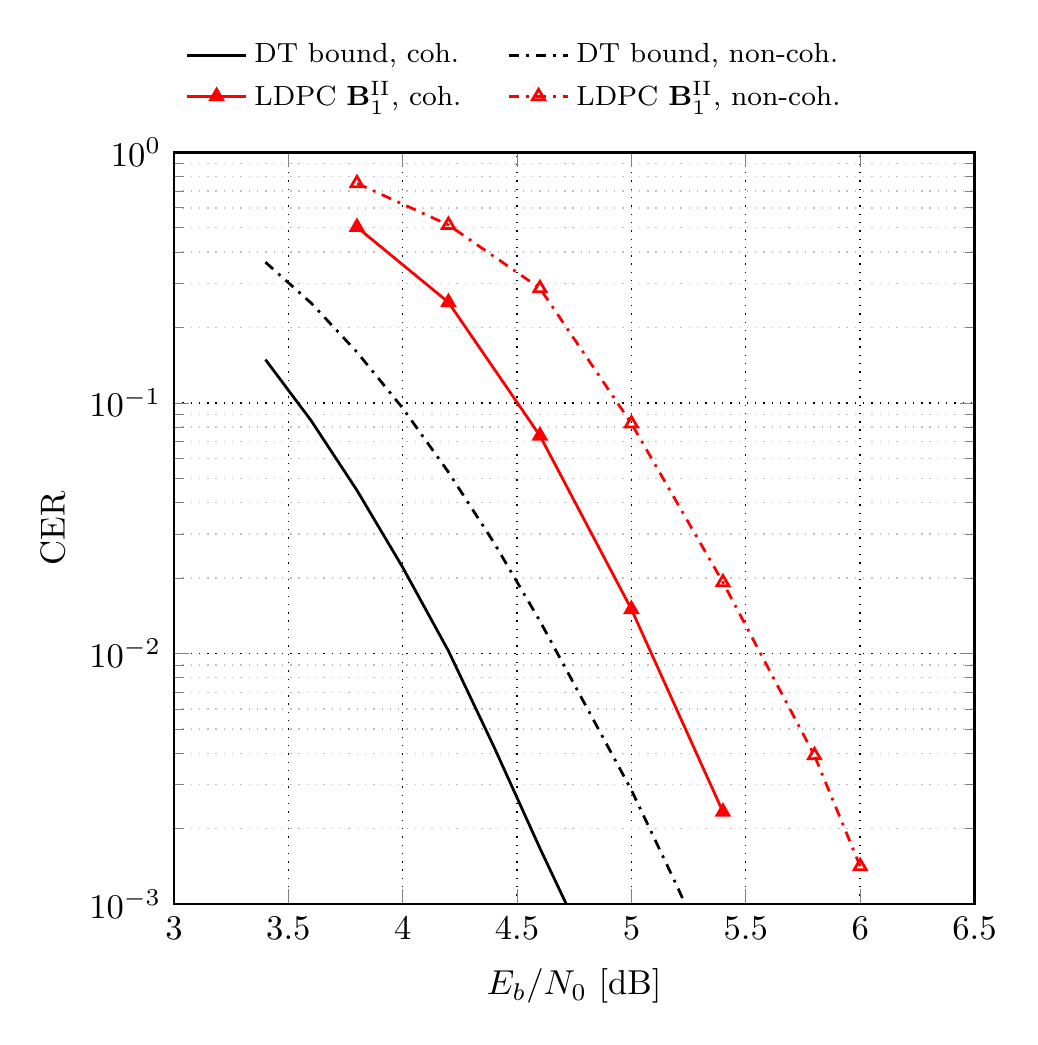}
	\caption{Simulation results of {an \ac{LDPC}} code over $\gf{8}$ with base matrix $\basem^\mathrm{II}_1$ combined with 8-\ac{DPSK} modulation, code rate $2/3$, $N=120$ symbols and phase-noise with $\sigma_{\Delta} = 2^\circ$ for the non-coherent case.}
	\label{fig:8psk_0.66}
\end{figure}

\begin{example}[$R_c=3/4$, $16$-\ac{DPSK}]
	Step 1 of the protograph search yields a set $\mathcal M_{e}$ of $1 \times 4$ base matrices. Its elements with the corresponding iterative decoding thresholds are given in Table~\ref{tab:thresh:16psk_r075}. {The Shannon limit for the coherent case is $\shLimCoh = 6.73$~ \textup dB. For the  non-coherent channel $\infoRateNonCoh = 7.12$~ \textup dB.} We find that the base matrix $\basem^{\mathrm{III}}_1=[2~2~2~1]$ has the best threshold $\threshNonCoh=8.24$~dB for the non-coherent phase noise channel. {We} {observe from the table that an ultra-sparse \ac{LDPC} code with regular \ac{VN} degrees of two would have $0.5$~dB worse threshold.} %We further have $\infoRateNonCoh = 7.12$~dB. 
	A rate-$3/4$ $(128,96)$ \ac{LDPC} code is obtained from it and its \ac{CER} versus $E_b/N_0$ curve is depicted in \fig{fig:16psk_0.75}\TN{,} together with the \ac{DT} bound for the coherent and non-coherent \ac{AWGN} channel. %To improve the waterfall performance, step 2b) in \secref{sec:prot_search} yields $\bar{\basem}^\mathrm{III}_1$. A code derived from this base matrix does not show a visible error floor.
	{We} observe from the figure a gap {with respect to} the \ac{DT} bound of around $1.2$~dB. 
\end{example}
\begin{table}[t]
	\centering
	\renewcommand{\arraystretch}{1.5}
	\caption{Iterative decoding thresholds for the non-coherent and coherent \ac{AWGN} channel for $16$-\ac{DPSK} modulation and rate-$3/4$ protographs.}	
	\label{tab:thresh:16psk_r075}
		\begin{tabular}{c|c|c|c}  
		\toprule
		&{Base Matrix} & $\threshNonCoh$ [dB] & $\threshCoh$ [dB] \\
		\midrule
		$\mathbf B^{\mathrm{III}}_1$ & $[2\ 2\ 2\ 1]$ &8.24& 7.85\\			
		$\mathbf B^{\mathrm{III}}_2$ & $[3\ 2\ 2\ 1]$ &8.57& 8.16\\	
		$\mathbf B^{\mathrm{III}}_3$ & $[3\ 3\ 2\ 1]$ &8.95& 8.50\\	
		$\mathbf B^{\mathrm{III}}_4$ & $[3\ 3\ 3\ 1]$ &9.31& 8.83 \\	
		$\mathbf B^{\mathrm{III}}_5$ & $[2\ 2\ 2\ 2]$ &8.76& 8.33\\	
		$\mathbf B^{\mathrm{III}}_6$ & $[2\ 2\ 2\ 3]$ &9.08&8.62 \\	
		$\mathbf B^{\mathrm{III}}_7$ & $[3\ 3\ 2\ 2]$ &9.40&8.93 \\	
		$\mathbf B^{\mathrm{III}}_8$ & $[3\ 3\ 3\ 2]$ &9.71&9.19 \\	
		$\mathbf B^{\mathrm{III}}_9$ & $[3\ 3\ 3\ 3]$ &9.97&9.45 \\	
		\bottomrule 					
	\end{tabular}
\end{table}

\begin{figure}[t]
	\centering	
	\includegraphics[scale=0.8, draft=false]{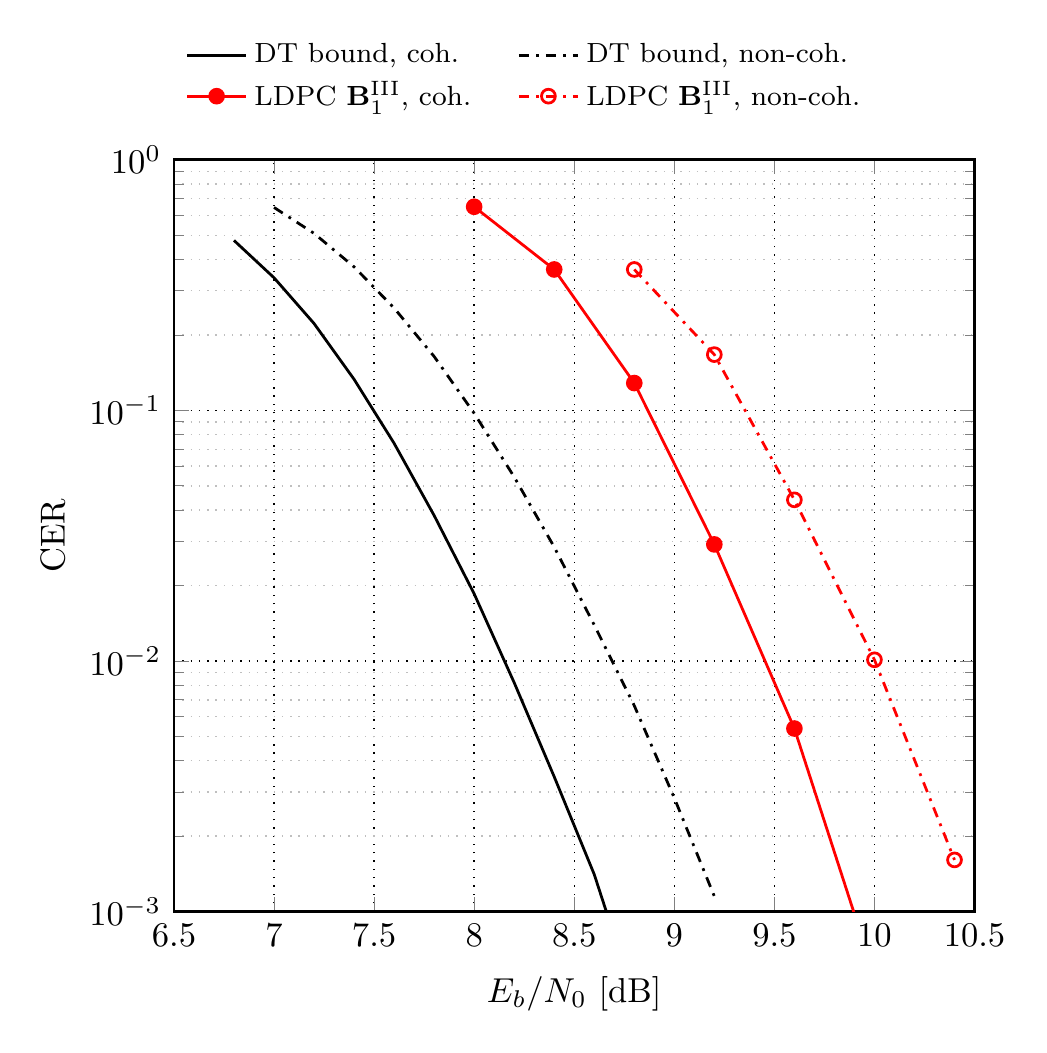}
	\caption{Simulation results {of an \ac{LDPC}} code over $\gf{16}$ with base matrix $\basem^\mathrm{III}_1$ combined with 16-\ac{PSK} modulation, code rate $3/4$, $N=128$ symbols and phase-noise having $\sigma_{\Delta} = 1^\circ$ for the non-coherent case.}
	\label{fig:16psk_0.75}
\end{figure}

\section{Conclusions}
\label{sec:conclusions}
In this work, we investigate the design of non-binary protograph \ac{LDPC} codes for the Wiener phase noise channel. We consider the serial concatenation of an outer $m$-ary \ac{LDPC} code over the finite field of order $m$ and  $m$-\ac{DPSK}\TN{,} and target transmission of short blocks in \TN{the} order of a few hundred symbols. Decoding of the concatenated scheme is {performed} in a turbo-like fashion where a detector and decoder iteratively exchange beliefs among each other.  We give a finite-length benchmark, namely the \ac{DT} bound, both for the coherent and non-coherent case. We show that, {with} a proper protograph \ac{LDPC} code design, a performance of $1.2$~dB or less from the \ac{DT} bound is achieved down to a \ac{CER} of $10^{-3}$, even in the presence of strong phase noise. All our designs are robust with respect to phase noise, in the sense that they nearly show the same gap to the respective bounds for both the coherent and non-coherent setup. {Furthermore, we observe that the protographs obtained for the Wiener phase noise channel are also the ones which have the best thresholds among all investigated protographs on the coherent channel.}

\bibliographystyle{IEEEtran}
\bibliography{./IEEEabrv_v3,./tn_bibl}

%\input{./biographies}
% use section* for acknowledgment
%\section*{Acknowledgment}

%The authors would like to thank...

% biography section
% 
% If you have an EPS/PDF photo (graphicx package needed) extra braces are
% needed around the contents of the optional argument to biography to prevent
% the LaTeX parser from getting confused when it sees the complicated
% \includegraphics command within an optional argument. (You could create
% your own custom macro containing the \includegraphics command to make things
% simpler here.)
%\begin{IEEEbiography}[{\includegraphics[width=1in,height=1.25in,clip,keepaspectratio]{mshell}}]{Michael Shell}
% or if you just want to reserve a space for a photo:

%\begin{IEEEbiography}{Michael Shell}
%Biography text here.
%\end{IEEEbiography}

% if you will not have a photo at all:
%\begin{IEEEbiographynophoto}{John Doe}
%Biography text here.
%\end{IEEEbiographynophoto}

% insert where needed to balance the two columns on the last page with
% biographies
%\newpage

%\begin{IEEEbiographynophoto}{Jane Doe}
%Biography text here.
%\end{IEEEbiographynophoto}

% You can push biographies down or up by placing
% a \vfill before or after them. The appropriate
% use of \vfill depends on what kind of text is
% on the last page and whether or not the columns
% are being equalized.

%\vfill

% Can be used to pull up biographies so that the bottom of the last one
% is flush with the other column.
%\enlargethispage{-5in}

% that's all folks
\end{document}